# System Efficiency vs. Individual Performance in Competing Systems

Chengling Gou

Physics Department, Beijing University of Aeronautics and Astronautics
37 Xueyuan Road, Haidian District, Beijing, China, 100083

Physics Department, University of Oxford
Clarendon Laboratory, Parks Road, Oxford, OX1 3PU, UK
gouchengling@hotmail.com, c.gou1@physics.ox.ac.uk

Abstract: this paper addresses the issue of the relation between the system efficiency and the individual performance with different combinations of agent memory lengths in mix-game model which is an extension of minority game (MG). In mix-game, there are two groups of agents; group1 plays the majority game, but the group2 plays the minority game. The average winnings of agents can represent the average individual performance and the volatility of a system can represent the efficiency of the system. It is found the correlations between the average winnings of agents and the medians of local volatilities are different when agent memory lengths change with different combinations of $m_1=m_2$, $m_1<m_2=6$ and $m_2<m_1=6$. This paper also gives some suggestions for designing complex competing systems.

Keywords: minority game, mix-game, system efficiency, individual performance

## 1. Introduction

A designer or a manager who design or manage competing systems which consist of heterogeneous agents who compete constantly for limited resources, such as different species in an ecological system, traders in financial markets, drivers on city roads and highways or users accessing a computer network – faces problems of coordination since selfish agents in such systems all want to optimize their performances and individual goals will most certainly be conflicting. If he or she hopes to balance the individual performance and the efficiency of the system, he or she needs to know the relationship between the individual performance and the efficiency of the system. However, the interplay between the state of optimal individual performance and the reach of a globally efficient phase where the collective use of resources is optimal may be quite subtle [1, 2]. Therefore, the study of this issue has both theoretical and practical importance, which can reveal how the macroscopic properties of a system may emerge from the microscopic interaction of agents and under which conditions, and this understanding may help for designing or managing of complex systems.

Agent-based models can be used to address directly the relation between microscopic behavior of agents and macroscopic properties of a system (like fluctuations, predictability and efficiency) [1, 2]. In article [2], I studied the correlations of the average winnings of agents and the volatilities of systems based on mix-game model [3, 4, 5] which is an extension of minority game (MG) [6, 7, 8]. In mix-game, there are two groups of agents; the group1 plays the majority game [9, 10, 11, 12, 13, 14 15, 16], but the group2 plays the minority game. The results show that the correlations between the average winnings of agents and the mean of local volatilities are different with different combinations of agents' memory lengths when the proportion of agents in group1 increases. The average winnings of agents can represent the average individual performance and the volatility of a system can represent the efficiency of the system. Therefore, these results imply that the combinations of agents' memory lengths largely influence the relation between the system efficiency and the average individual performance.

In this paper, I further address this issue by examining how the combinations of agents' memory lengths influence the relation between the system efficiency and the average individual performance. This paper is organized as following. Section 2 describes the mix-game model and the simulation conditions. Section 3 presents the simulation results and the results of the quantitative correlations. Section 4 discusses the results. In section 5, the conclusion is reached.

2. **The model and simulation conditions**

Mix-game model is an extension of minority game (MG) so its structure is similar to MG. In mix-game, there are two groups of agents; the group1 plays the majority game, and the group2 plays the minority game. N (odd number) is the total number of the agents and N1 is number of agents in group1. The system resource is $r=N*L$, where $L<1$ is the proportion of resource of the system. All agents compete in the system for the limited resource r. T1 and T2 are the time horizon lengths of the two groups of agents, and m1 and m2 denote the memory lengths of the two groups of agents, respectively.

The global information only available to the agents is a common bit-string "memory" of the m1 or m2 most recent competition outcomes (1 or 0). A strategy consists of a response, i.e., 0 (sell) or 1 (buy), to each possible bit string; hence there are $2^{2^{m1}}$ or $2^{2^{m2}}$ possible strategies for group 1 or group 2, respectively, which form full strategy spaces (FSS). At the beginning of the game, each agent is assigned s strategies and keeps them unchangeable during the game. After each turn, agents assign one (virtual) point to a strategy which would have predicted the correct outcome. For agents in group 1, they will reward their strategies one point if they are in the majority; for agents in group 2, they will reward their strategies one point if they are in the minority. Agents collect the virtual points for their strategies over the time horizon T1 or T2, and they use their strategy which has the highest virtual point in each turn. If there are two strategies which have the highest virtual point, agents use coin toss to decide which strategy to be used. Excess demand is equal to the number of ones (buy) which agents choose minus the number of zeros (sell) which agents choose. According to a widely accepted assumption that excess demand exerts a force on the price of the asset and the change of price is proportion to the excess demand in a financial market [17, 18, 19], the time series of price of the asset can be calculated based on the time series of excess demand.

In simulation, the distribution of initial strategies of agents is randomly distributed in the full strategy space and keeps unchanged during the game. Simulation turns are 3000. Total number of agents is 201. Number of strategies per agent is 2.

### 3. Simulation results and discussion

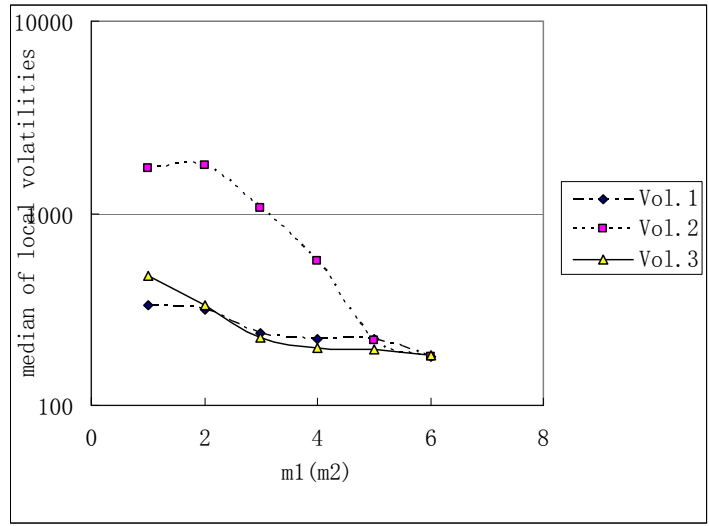

Fig.1 medians of local volatilities vs. different memory length of group1 or group 2 (m1 or m2), Vol.1 representing medians of local volatilities of m2=6, T1=T2=12, N=201, N1=72, s=2 and m1 varying from 1 to 6; Vol.2 representing medians of local volatilities of m1=6, T1=T2=12, N=201, N1=72, s=2 and m2 varying from 1 to 6; Vol.3 representing medians of local volatilities of m1=m2, T1=T2=12, N=201, N1=72, s=2.

Volatility of the time series of prices is represented by the variance of the increases of prices. The local volatilities are calculated at every timestep by calculating the variance of the increases of prices within a small timestep window. The window length of local volatility is 5 in Fig.1. From Fig.1, one can find that the medians of local volatilities (Vol.1 and Vol.2) increase while the memory length m1 or m2 decreases from 6 to 1 in both simulation situations. But the difference is that Vol.2 is much larger than Vol.1 when m1 or m2 is smaller than 5. This means that agents with larger memory length playing majority game make the systems less efficient. Vol.3 represents the medians of local volatilities when m1 is equal to m2 and they increase from 1 to 6 and it is similar to Vol.1.

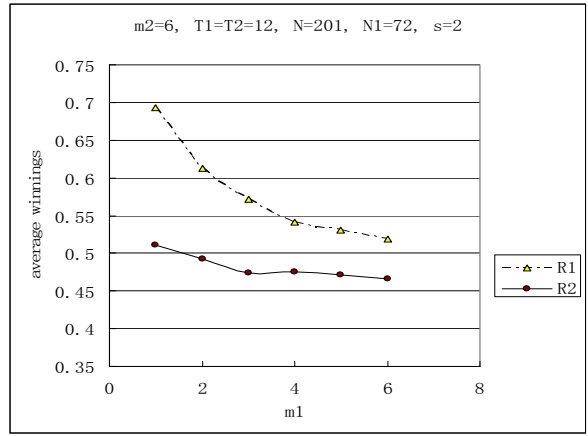

Fig.2 average winnings per agent per turn for Mix-game vs. different memory length of group1 when m2=6, T1=T2=12, N=201, N1=72 and s=2. R1 represents the average winning per agent per turn in group1, and R2 represents the average winning per agent per turn in group2.

Table 1 the correlations among R1, R2 and Vol.1 when m2=6, and m1 increases from 1 to 6

|       | R 1      | R 2      | Vol.1 |
|-------|----------|----------|-------|
| R 1   | 1        |          |       |
| R 2   | 0.982427 | 1        |       |
| Vol.1 | 0.935162 | 0.954744 | 1     |

From Fig. 2, one can find that the average winnings (R1 and R2) of these two groups increase when m1 decreases from 6 to 1 and R1 increases much more quickly than R2. The average winning of group1 (R1) is larger than that of group2 (R2). Comparing Fig.1 with Fig.2, one can find that the median of local volatilities (Vol.1) increases when m1 decreases from 6 to 1 under the condition of m2=6, T1=T2=12, N=201, N1=72, s=2, accompanying with the increase of the average winnings of group1 and group2 (R1, R2). Table 1 shows that correlations are among R1, R2 and Vol.1 are largely positive. These results mean that the improvement of the performance of individual agents accompanies with the decrease of the efficiency of systems under this simulation condition. Agents can benefit from the larger fluctuation of systems, especially for agents in group1. This is accordant with the reality of financial markets.

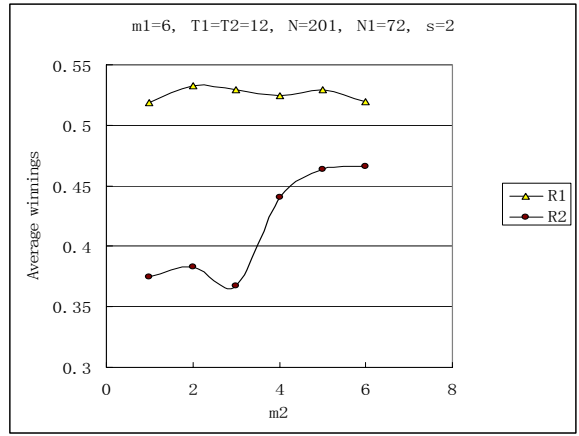

Fig.3 average winnings per agent per turn for Mix-game vs. different memory length of group2 when m1=6, T1=T2=12, N=201, N1=72 and s=2. R1, R2 have the same meaning as indicated in Fig.2.

Table 2 correlations among R1, R2 and Vol.2 when m1=6, and m2 increases from 1 to 6

| correlation | R1 | R2 | Vol.2 |
|---|---|---|---|
| R1 | 1 | | |
| R2 | -0.246 | 1 | |
| Vol.2 | 0.187 | -0.899 | 1 |

From Fig. 3, one can find that the average winning of group2 (R2) decreases when m2 decreases from 6 to 1 while the average winning of group1 (R1) does not change greatly. The average winning of group1 (R1) is much larger than that of group2 (R2) and the agents in group1 don't influence the average real winnings of agents in group2 in this simulation condition [5].

Comparing Fig.3 with Fig.1, one can notice that the median of local volatilities (Vol.2) increases when m2 decreases from 6 to 1 under the condition of m1=6, T1=T2=12, N=201, N1=72, s=2, accompanying with the decrease of the average winning of group2 (R2) and nearly stable average winning of group1 (R1). Table 2 shows that the average winning of group2 (R2) strongly negatively correlated with the volatilities of systems (Vol.2), but the average winning of group1 (R1) just slightly positively correlated with the volatilities of systems (Vol.2).

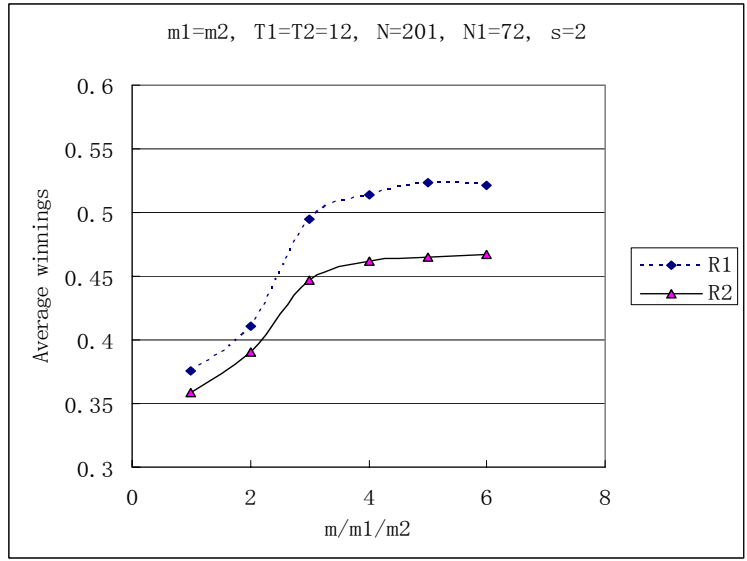

Fig.4 average winnings per agent per turn for Mix-game vs. different memory length of group2 when m1=m2, T1=T2=12, N=201, N1=72 and s=2. R1, R2 have the same meaning as indicated in Fig.2.

Table 3 correlations among R1, R2 and Vol.3 when m1=m2, and they increase from 1 to 6

|       | R1       | R2       | Vol.3 |
|-------|----------|----------|-------|
| R1    | 1        |          |       |
| R2    | 0.976722 | 1        |       |
| Vol.3 | -0.80357 | -0.66474 | 1     |

From Fig.4 one can find that the average winnings of group1 and group2 (R1 and R2) have the lowest points at m1=m2=2 and the highest points at m1=m2=4. Comparing Fig.4 with Fig.1, one can observe that the relations among that the average winnings of group1 and group2 (R1 and R2) and the volatilities of systems (Vol.3) are much more complicated than that in the previous two simulation situations. Comparing R1 and R2 with Vol.3 at m1=m2=1 and m1=m2=6, we can notice that agents with larger memory length get higher winning scores but the Vol.3 remains nearly the same. However, at m1=m2=3, Vol.3 has the highest value, but R1 at this point is just a little larger than that at m1=m2=1 and R2 is almost the same at these two points. It is interesting that R1 and R2 increase greatly at m1=m2=4 and Vol.3 decreases obviously at the same point. Table 3 shows that the average winnings of group1 and group2 (R1 and R2) greatly negatively correlated with the volatilities of systems

(Vol.3).

## 4. Concluding remarks

The correlations between the average winnings of agents (R1 and R2) and the medians of local volatilities are different when agents' memory lengths change with different combination of m1=m2, m1<m2=6 and m2<m1=6.

If a designer or a manager of a complex system hope to reach an optimal state at which the system has high efficiency and agents have high individual performances, he or she can choose m1=m2=4, 5 and 6 as agents' memory lengths, and avoid the memory length combinations of m1=6>m2. If he or she intend to improve the agents' performance at the cost of moderately losing system efficiency, he or she can m1<m2=6 as agents' memory lengths. This study result also suggests that m1 be smaller than m2 if researchers want to mix-game model to simulate the financial markets.

Acknowledgements

This research is supported by Chinese Scholarship Council. Thanks Professor Neil F. Johnson for discussion. Thanks David Smith for providing the original program code MG.